# Insulator to semi-metal transition in graphene oxide

Goki Eda, Cecilia Mattevi, Hisato Yamaguchi, HoKwon Kim and Manish Chhowalla*

*Rutgers University, Materials Science and Engineering, Piscataway, NJ 08854*

Transport properties of progressively reduced graphene oxide (GO) are described. Evolution of the electronic properties reveals that as-synthesized GO undergoes insulator-semiconductor-semi-metal transitions with reduction. The apparent transport gap ranges from 10 ~ 50 meV and approaches zero with extensive reduction. Measurements at varying degrees of reduction reveal that transport in reduced GO occurs via variable-range hopping and further reduction leads to increased number of available hopping sites.

* Corresponding author: manish1@rci.rutgers.edu

Graphene oxide (GO), a chemically derived graphene, has recently triggered research interest due to its solubility in a variety of solvents and promise of large area electronics[1-3]. In contrast to extensive studies on mechanically exfoliated graphene[4], the electronic transport properties of chemically derived graphene has received little attention due to its moderate properties. However, from a fundamental point of view, transport in GO is intriguing due to the presence of substantial electronic disorder arising from variable $sp^2$ and $sp^3$ bonds. GO can be viewed as graphene with oxygen functional groups decorating the basal plane and edges. The carbon-oxygen bonds in GO are $sp^3$ hybridized which disrupts the extended $sp^2$ conjugated network of graphene. Oxidation also generates various types of defects in the graphene lattice, which limits transport[5]. The substantial $sp^3$ fraction in GO (~ 0.6)[6] renders it insulating but incremental removal of oxygen can transform the material to a semiconductor and ultimately to a graphene-like semi-metal[7].

The electronic band structure of GO is not clearly understood due to the nano-scale inhomogeneities in the structure[8]. Recent optical studies[9, 10] indicate that GO possesses an energy gap which can be tuned by controlling the degree of oxidation or reduction. The possibility of band gap engineering in GO is of interest for its implementation in photonic and electronic devices. However, a complete study elucidating the transport mechanism in an individual GO sheet at different degrees of reduction has been lacking. Jung *et al.*[7] reported a gradual transition of GO from electrical insulator to graphene-like semi-metal with thermal reduction but did not provide insight into the how this occurs. Their results suggest an intimate relationship between the chemical structure and the electrical properties of GO. Kaiser *et al.*[11] recently conducted a detailed study of the transport through an individual GO sheet reduced by hydrogen plasma treatment. Their analysis indicated that transport in their reduced GO (r-GO) occurs via variable range hopping between localized $sp^2$ states. The results in Ref. [11], however, are limited to GO reduced at the optimum level. Since the electrical properties of r-GO are strongly correlated to the amount of $sp^3$ bonding (i.e. the amount of residual oxygen), which represent transport barriers[6], additional studies correlating the transport properties at different degrees of reduction would be useful for fundamental insight into the insulator-semi-metal transition.

We have extended works of Jung *et al.*[7] and Kaiser *et al.*[11] to provide a complete picture of transport in GO by carrying out measurements over a wide temperature range at various degrees of reduction treatments. Specifically, we report the temperature dependent transport properties of individual GO sheet field effect devices as a function of progressive reduction treatment. We have observed a transport gap at low temperatures for moderately reduced GO. Further, we find that the energy gap is dependent on the extent of the reduction process and approaches zero with extensive reduction. Our results suggest that conduction in r-GO is limited by defects and occurs from a combination of hopping and thermal activation at mid-gap states, as is commonly observed in non-crystalline materials[12].

Graphene oxide was prepared using a modified Hummers method[13] and deposited on a $SiO_2$ (300 nm)/Si substrates with pre-patterned alignment marks. The positions of the monolayer GO flakes with respect to the alignment marks were identified under the optical microscope and conventional electron-beam lithography was used to define electrodes on the sheets. The GO sheets were contacted by thermally evaporating Cr/Au (5/30 nm) followed by a lift-off process. Multiple electrodes with varying separations were patterned on individual flakes in order to study the effect of contact resistance. Reduction of GO was achieved by exposure to saturated vapor of hydrazine monohydrate in a loosely sealed petri dish at 80 $^{o}$C over variable length of time. All electrical measurements were conducted on the same device after each step of reduction. In the following discussions, the samples referred to as 8m, 15m, 30m, 16h are identical GO devices with 8, 15, 30 minutes, and 16 hours of exposure to hydrazine, respectively. In addition to the GO devices reduced only after deposition, several GO flakes reduced directly in anhydrous hydrazine prior to deposition were also studied. It has been reported that solubilization of GO directly in anhydrous hydrazine leads to the most efficient reduction which leads to dramatic improvements in transport[14]. For optimum reduction and removal of residual hydrazine, these devices were annealed in $N_2/H_2$ (90/10) atmosphere at 150 $^{o}$C for 1 hour prior to measurements. Results from two such hydrazine dipped r-GO samples are presented in the following discussions and are referred to as HG-A and HG-B. The transport properties of r-GO devices were studied in two-terminal configurations with silicon substrate used to apply gate bias $V_g$. All measurements were made in vacuum at temperatures ranging from 78 to 360 K.

The room-temperature conductivity of r-GO devices varied over 5 orders of magnitude depending on the degree of reduction with 8m and HG-A sample exhibiting the lowest and the highest conductivity, respectively. The transfer characteristics of 6 different r-GO devices at $T$ = 300 K and $T$ = 78 K are shown in Fig 1. Ambipolar field effect with minimum conductivity around $V_g \sim 0$ V was observed for all devices, indicating that the doping effect is minimal. While none of the devices exhibit insulating state at $T$ = 300 K, all devices exhibit improved on-off behavior at low temperatures. The distinct off-state observed for samples 8m, 15m, and 30m at $T$ = 78 K suggest that the energy gap is larger for lightly reduced GO compared to extensively reduced samples such as HG-A and HG-B. The on/off ratios achieved in 15m and 30m samples exceed $10^3$, which is more than an order of magnitude greater than those reported for lithographically patterned GNRs measured at comparable temperatures[15]. This comparison suggests that the size of the band gaps in these r-GO are comparable to or greater than those of GNRs, which are of the order of few tens to few hundreds of milli-electron volts depending on their width[15].

One conduction mechanism proposed for GO and r-GO is Schottky emission, where a Schottky barrier near the electrode interface limits charge injection[16, 17]. To study the effect of these interfaces, the *I*-*V* characteristics of r-GO were studied as a function of channel length $L$ as shown in Fig 2a. Apparent potential profiles across the r-GO were obtained by plotting the bias voltage required to achieve an arbitrary width-normalized current as a function of $L$, as shown in Fig 2b for different gate bias conditions. This analysis shows that the voltage drop across r-GO is nearly linear and approaches zero for $L \to 0$ independent of the $V_g$, indicating that the contribution from the electrode interface is minimal and the conduction is bulk limited. The principle mechanism of our r-GO devices are therefore not Schottky barrier modulation as commonly seen in single-walled carbon nanotube FETs[18].

Kaiser *et al.*[11] reported that electrical conduction in r-GO can be explained by variable-range hopping (VRH) between the localized states. In VRH model, the temperature dependence of the conductivity $\sigma$ can be described by the form

$$\sigma = A \exp\left(-\frac{B}{T^{1/3}}\right). \tag{1}$$

The parameters $A$ and $B$ are expressed as

$$A = \frac{eR_0^2 \nu_{\mathrm{ph}}}{k_{\mathrm{B}}}$$

$$B = \left(\frac{3\alpha^2}{N(E_{\mathrm{F}})k_{\mathrm{B}}}\right)^{1/3}$$

where $e$ is the electronic charge, $R_0$ is the optimum hopping distance, $\nu_{\mathrm{ph}}$ is the frequency of the phonons associated with the hopping process, $k_{\mathrm{B}}$ is the Boltzmann constant, $\alpha$ is the wavefunction decay constant, and $N(E_{\mathrm{F}})$ is the density of states near the Fermi level. Here, we examine the minimum conductivity $\sigma_{\mathrm{min}}$ of r-GO instead of $\sigma(V_{\mathrm{g}} = 0\ \mathrm{V})$ in order to exclude the effects of charged impurities[19]. Fig 3a is a plot of $\sigma_{\mathrm{min}}$ as a function of $T^{-1/3}$ showing reasonable agreement with the VRH model. It can be noticed that the linear trend continues up to room temperature for lightly reduced GO devices (such as 15m and 30m) whereas deviation from the VRH model is observed above ~ 240 K for well-reduced GO devices (16h, HG-A and HG-B). The higher temperature regime of well-reduced GO can be fitted reasonably well with the Arrhenius model, suggesting that thermally excited carriers begin to dominate electrical conduction. The temperature at which the cross-over occurs (indicated by arrows in Fig 3a) decreases with the extent of reduction, which is in agreement with the prediction that restoration of percolating $sp^2$ carbon network allows band-like transport[20]. Indeed, Chua *et al.*[21] demonstrated that the Arrhenius type conduction is more likely to occur for r-GO with minimal oxidation. The deviation from the VRH behavior for 15m sample at low temperatures requires more analysis but we believe that it is related to the presence of deep trap states, which prevent the precise determination of $\sigma_{\mathrm{min}}$.

Additional evidence supporting hopping conduction can be observed in the average field ($F = V_{\mathrm{sd}}/L$) dependence of the field effect mobility $\mu$ measured at room temperature (Fig 3b). The field effect mobility was obtained from the linear region of $\sigma$ vs $V_{\mathrm{g}}$ plot assuming $\sigma(V_{\mathrm{g}})^{-1} = \sigma_{\mathrm{min}}^{-1} + (C_{\mathrm{ox}}\ (V_{\mathrm{g}} - V_{\mathrm{th}})\mu)^{-1}$ where $C_{\mathrm{ox}}$ is the oxide capacitance and $V_{\mathrm{th}}$ is the threshold voltage at which charge neutrality condition is reached. The field dependence is nearly exponential, in reasonable agreement with the calculations[22]. Interestingly, the mobility of HG-B sample is independent of $F$, which is consistent with the fact that band-like transport is expected for this sample at room temperature. This finding suggests that the trap levels for well reduced GO lie close the mobility edge such that thermal excitation is sufficient to de-trap the carriers.

Our previous studies indicate that as-prepared GO has a semi-amorphous structure with moderate long-range order[8] which is expected to give rise to band tail states. In addition, $sp^2$ carbon clusters and filaments surrounded physical barriers represented by $sp^3$ carbon atoms give rise to localized states. In order to gain insight into the energies associated with the electronic structure of r-GO, the temperature dependence of the intrinsic carrier concentration $n_i$ was investigated. At the charge neutrality condition, there is equal number of carriers with opposite charge and $\rho_{max} = (\sigma_{min})^{-1} = (en_i(\mu_h + \mu_e))^{-1}$ where $\mu_h$ and $\mu_e$ are mobilities of holes and electrons, respectively. It should be noted that both $n_i$ and $\mu$ are temperature dependent quantities for r-GO, in contrast to mechanically exfoliated graphene[23, 24]. The Arrhenius plot of $n_i$ shows at least 2 regimes with different slopes, suggesting that the density of states is non-monotonous (Fig 4a). The activation energy $E_a$ extracted from the low and high temperature regimes are designated as $E_{a,1}$ and $E_{a,2}$, respectively. The relationship between the energy levels and transitions associated with $E_{a,1}$ and $E_{a,2}$ is not trivial. However, based on the fact that small energy gap on the order of the thermal energy is present for lightly reduced GO samples, it is reasonable to assume that $E_{a,1}$, which is found to be equal to or less than 55 meV for all samples, as the apparent energy gap between the tail states of the valence and the conduction bands. On the other hand, $E_{a,2}$, which is larger than $E_{a,1}$ by a factor of about 3~8, may be associated with transitions between energy states away from $E_F$. Fig 4b shows $E_{a,1}$ and $E_{a,2}$ plotted as a function of the maximum resistivity measured at room temperature $\rho_{max}^{RT}$, which is a measure of the degree of reduction. It is clearly shown that both $E_{a,1}$ and $E_{a,2}$ decrease with $\rho_{max}^{RT}$ and approaches zero, indicating narrowing and closing of the energy gap with reduction.

Further insight into the electronic structure of r-GO is gained by investigating the parameter $B$ of Eqn (1). The general form of VRH described by Eqn (1) assumes that the Fermi energy lies in the range of the localized states and that the density of states is constant within several $k_B T$ near the Fermi level. At temperatures approaching 0 K, we can assume $\partial n_i / \partial T = N(E_F)k_B$. We estimated this quantity by taking $n_i$(100 K)/(100 K). Although this is a crude approximation, the value of $\partial n_i / \partial T$ obtained thus provide a reasonable extrapolation to $n_i$ vs $T$ plot in the low temperature region. Fig 4c shows $N(E_F)$ and localization length $1/\alpha$ plotted as a function of $\rho_{max}^{RT}$. This analysis leads to an interesting result which is that reduction of GO does not lead to delocalization of carriers but to increased number of localized states near $E_F$ at low temperatures. Although counterintuitive, this result is consistent with the fact that despite extensive reduction, the coherence length $L_c$ of r-GO obtained via Raman spectroscopy remains nearly constant[25]. We also remark that the values of $1/\alpha$ is surprisingly in good agreement with $L_c$ obtained in our previous study[6].

In summary, the transport properties of r-GO as a function of progressive reduction treatment have been investigated. The carrier transport in lightly reduced GO was shown to occur via variable-range hopping whereas band-like transport begins to dominate in well-reduced GO. The energy gap of between the tail states of valence and conduction bands is on the order of 10 ~ 50 meV which is significantly smaller than the reported optical gap of as-synthesized GO[9, 10]. We further demonstrate that reduction of GO leads to increased number of localized states while the localization length remains largely unchanged.

Figure Captions

Fig 1: Transfer characteristics of r-GO with different degree of reduction measured at $T$ = 300 K and $T$ = 78 K. The inset shows an optical micrograph of a typical r-GO device.

Fig 2: (a) Width-normalized current ($I_{sd}$/W) as a function of source-drain bias ($V_{sd}$) from a single r-GO flake (15m sample) measured at different probe separations ($L$) shown in μm. (b) Apparent potential profile across a r-GO flake (15m sample) at different gate bias ($V_g$) extracted from analysis of length-dependent $I_{sd}$/W - $V_{sd}$ characteristics. See text for details.

Fig 3: (a) Minimum conductivity $\sigma_{min}$ of r-GO as a function of $T^{-1/3}$. The linear fits show agreement with the VRH transport. For samples 16h, HG-A, and HG-B, deviation to thermally activated transport is observed at temperatures indicated by the arrows. (b) Field effect mobility $\mu$ as a function of average electric field $F$. Open and filled symbols corresponds to hole and electron mobility, respectively.

Fig 4: (a) Arrhenius plot of intrinsic carrier mobility $n_i$. Two regions are fitted with a line with different slopes. The inset is a schematic illustration of the energy band structure of r-GO showing localized states near $E_F$ and band tails. (b) Thermal activation energies ($E_{a,1}$ and $E_{a,2}$) as a function of the room temperature maximum resistivity $\rho_{max}^{RT}$ of r-GO. (c) Density of states at the Fermi level $N(E_F)$ and localization length $1/\alpha$ as function of $\rho_{max}^{RT}$.

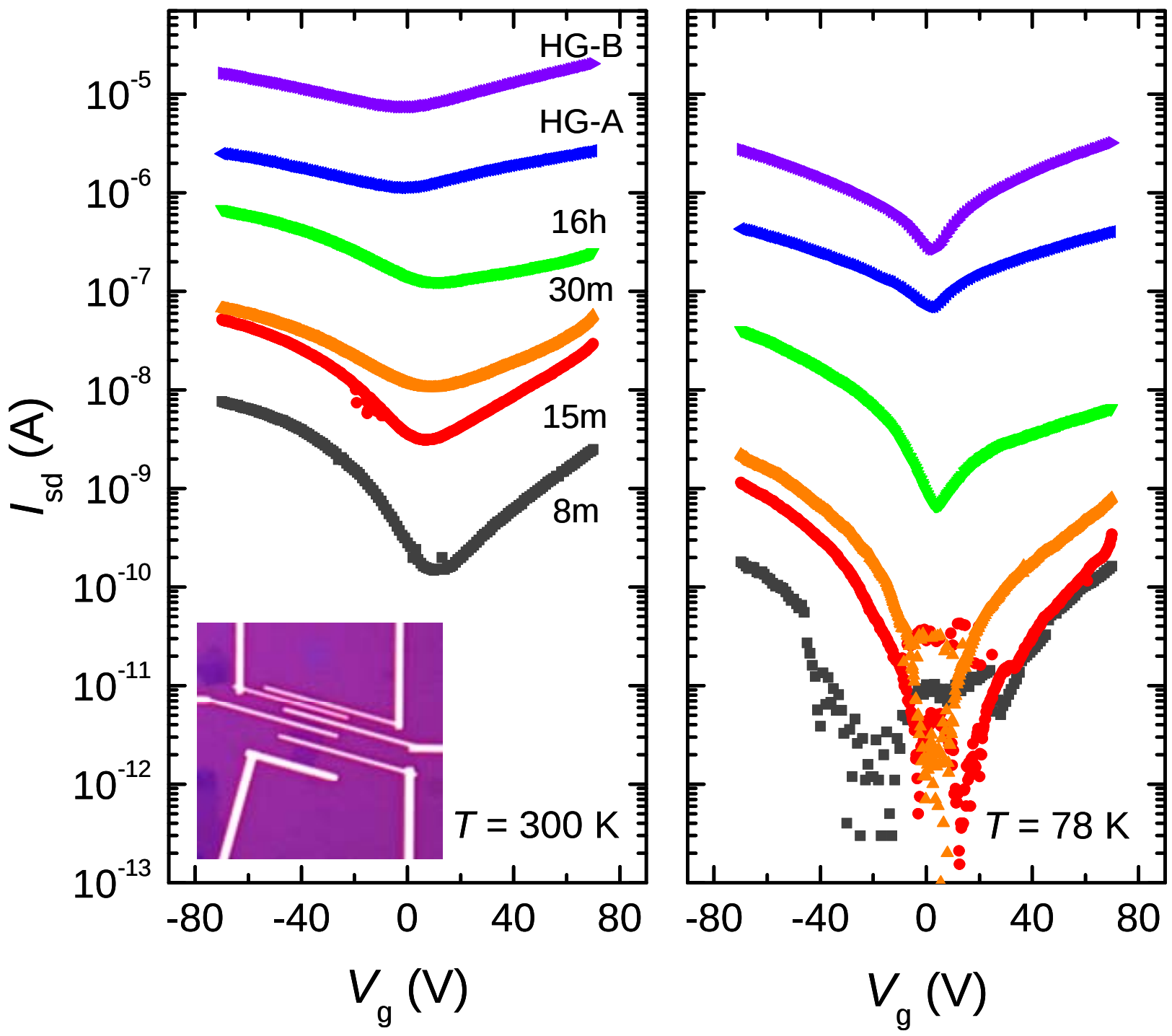

Fig 1

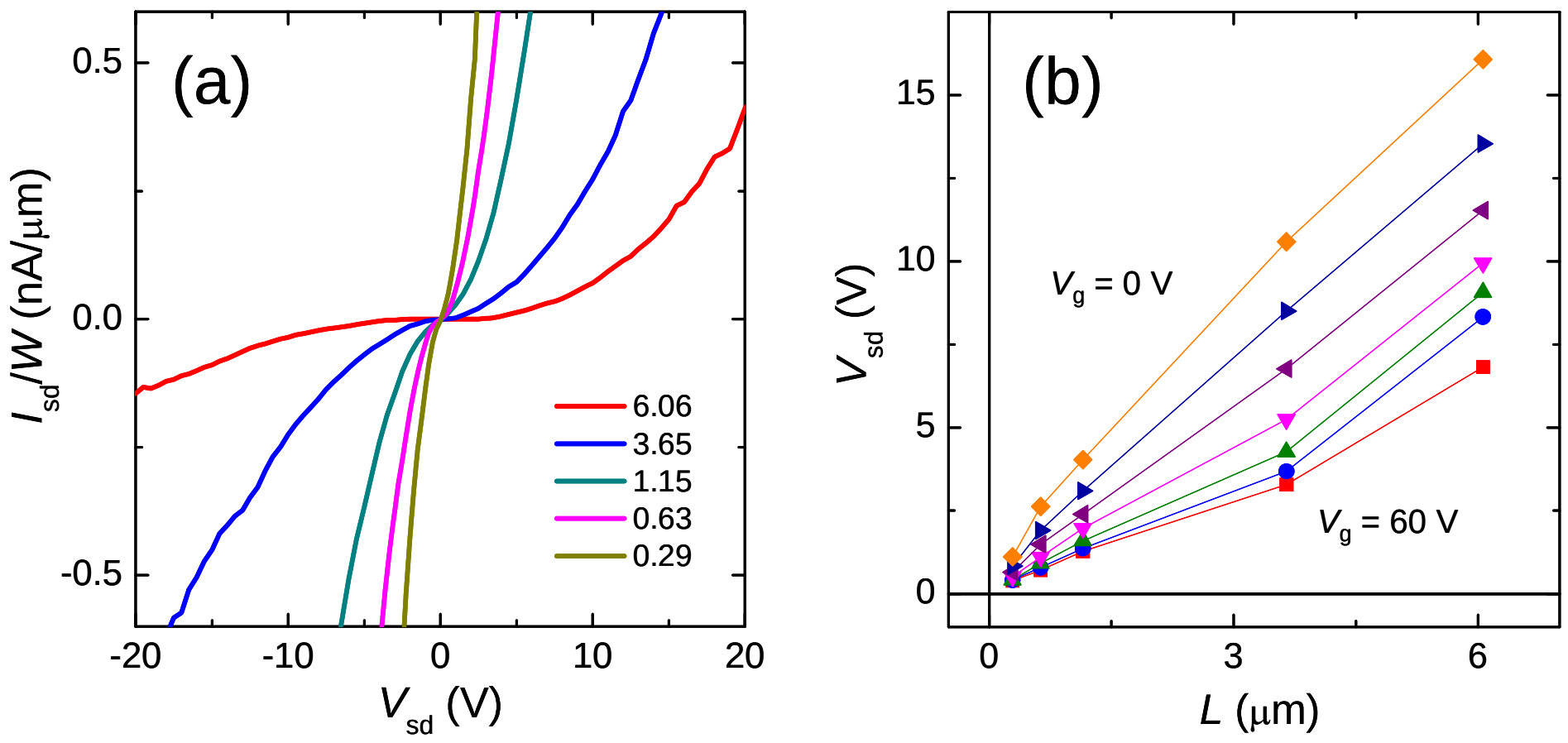

Fig 2

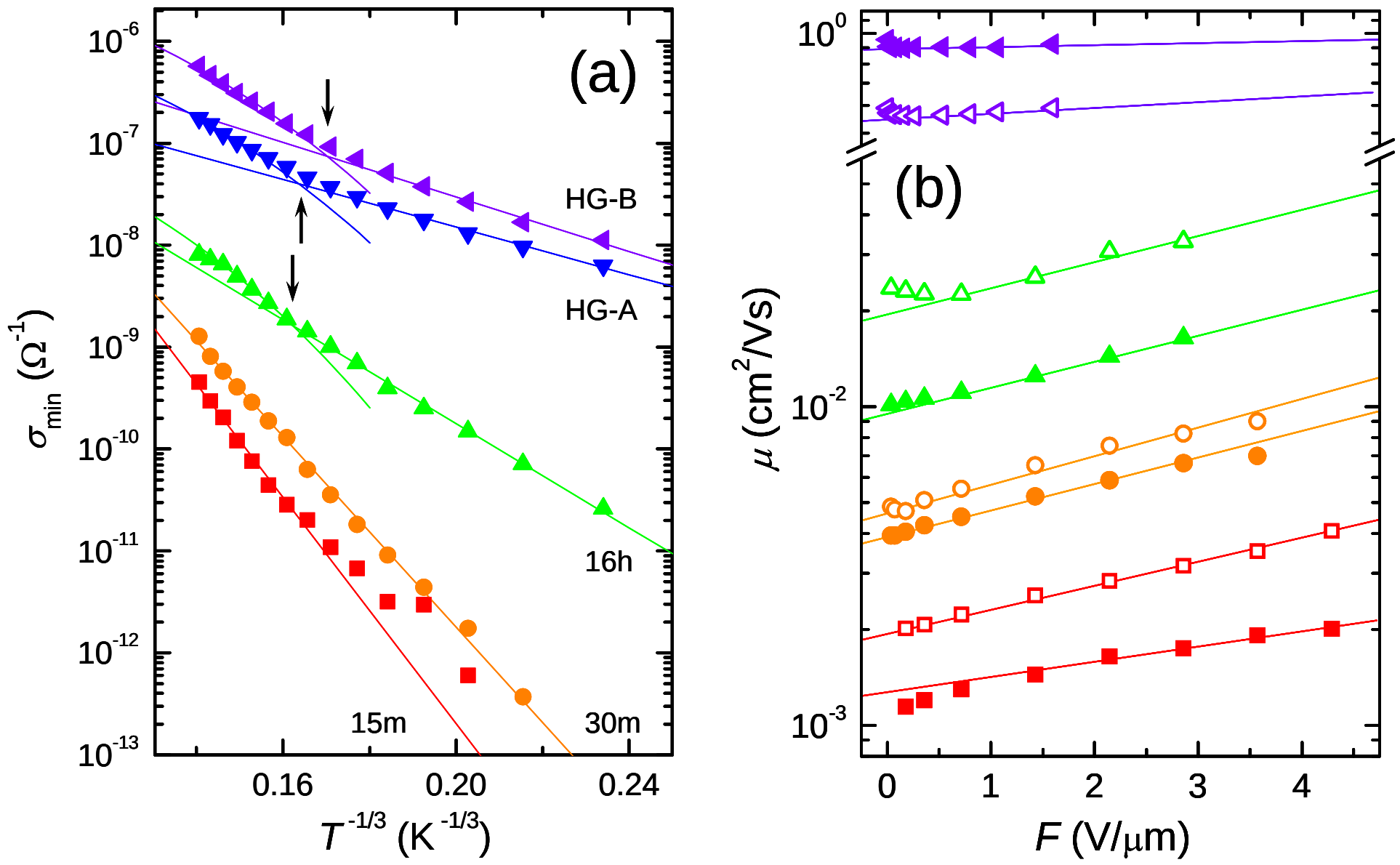

Fig 3

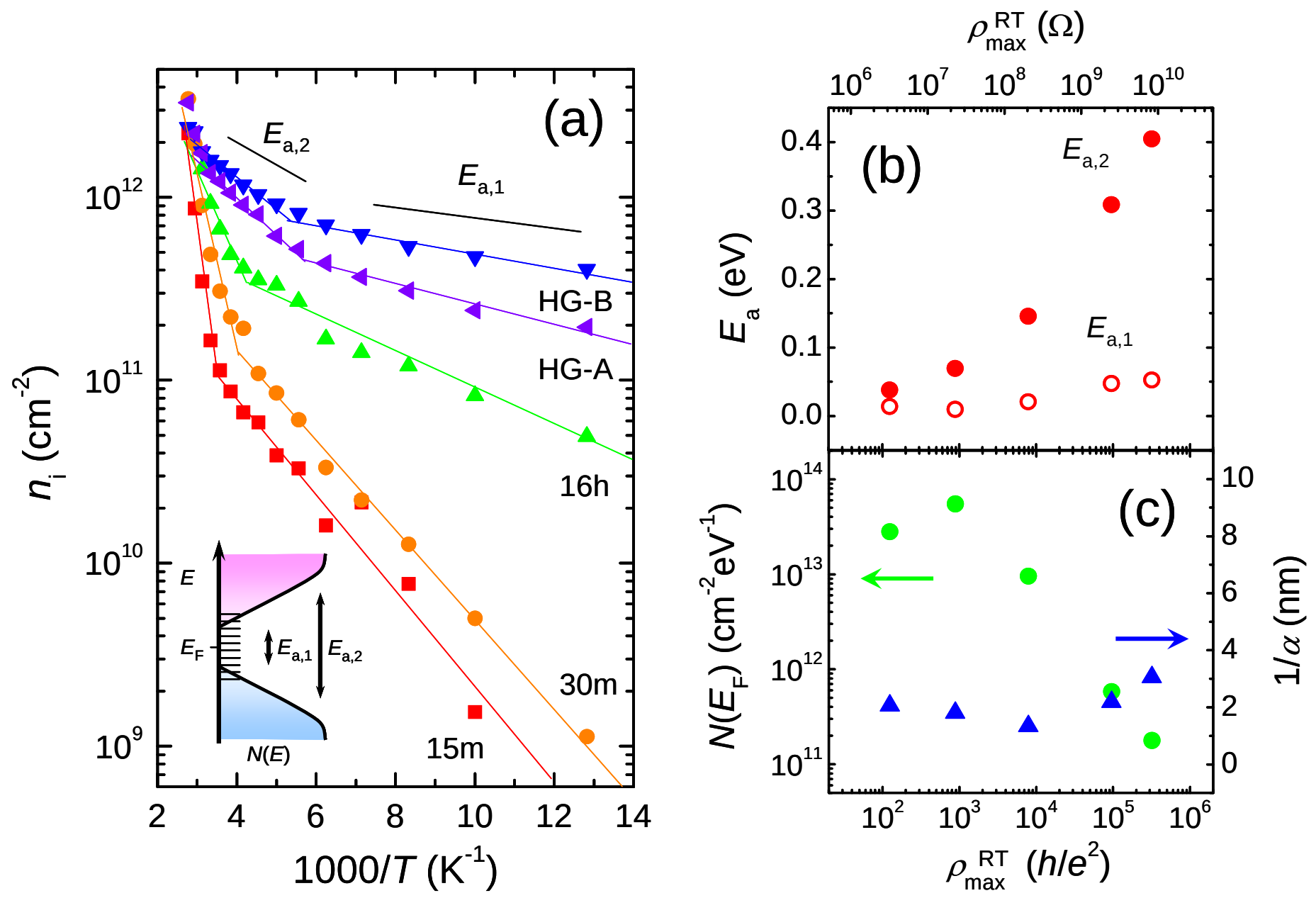

Fig 4